          [2001/04/25 1.1 (PWD)]
\documentclass[a4paper,twocolumn]{esapub2005} 
\usepackage{times}
\usepackage{natbib}
\usepackage{graphicx}

\title{Observing the high energy behaviour of the low mass X-ray binary XB 1832--330 with {\it INTEGRAL\/}}
\author{A. Tarana}
\affil{IASF-INAF Rome, Italy; Universit\`a Tor Vergata di Roma, Italy}
\author{A. Bazzano}
\author{P. Ubertini}
\author{M. Federici}
\affil{IASF-INAF Rome, Italy}

\newcommand{\source}{XB 1832--330}
\begin{document}


\maketitle

\begin{abstract}
We  report {\it INTEGRAL\/}/IBIS results on temporal and spectral behaviour of the  Low Mass X-ray Binary \source\ located in the globular cluster NGC 6652. During the 2003-2005 monitoring of the Galactic Centre, the source shows a weak flux variability and an everage flux in the 20--150 keV of about  2$\times 10^{-10}$ erg s$^{-1}$ cm$^{-2}$. The overall energy spectrum extends up to 150 keV  and is well described by a Comptonization model with an electrons temperature of $\sim$ 22 keV and a plasma optical depth of  1.8. We thus confirm the persistent nature of this burtser. 

\end{abstract}

\section{Introduction}

Firstly reported as H 1825--33 from observation with the Einstein observatory (HEAO-1) \citep{hw}, and later confirmed by ROSAT as a X-ray source member of the Galactic globular cluster NGC 6652 \citep{phv}, this source was then catalogated as atoll type Low Mass X-ray Binary under XB 1832--330 \citep{v}.

The long exposure on the galactic bulge region performed with the WFCs on board BeppoSAX allowed to discover the precence of Type I X-ray bursts emission \citep{i}. This signature is the evidence of the neutron star nature of the compact object.

A broad band (0.1--200 keV) spectral study of the persistent emission was made by Parmar et al. (2001) with the NFs Instrument of the BeppoSAX satellite. The spectrum was well  described by a disk-blackbody and a comptonization (CompTT) component plus the addition of a partial covering of the disk. A 0.1-200 keV luminosity of 4.4 $\times 10^{36}$ erg s$^{-1}$ has been reported for a distance of 9.2 kpc (derived from optical method \citep{c}).

This system belong to the globular cluster X-ray sources in binary systems characterized by short period ($<$ 1 hr) and its optical counterpart is identified with a blue variable object with $M_ {V}=3.7$ \citep{heike}.

\section{Data analisys}

Data analysis includes all the public data in which XB 1832-330 was within the ISGRI/IBIS fully coded field of view ($9^{\circ} \times$9$^{\circ}$) \citep{u} so that the source intensity determination is not effected by possible systematic errors due to uncertainties in the off-axis responce. The observations cover the period from 2003 March 22 to 2005 September 25 for a total of 132 SCWs (pointing lasting about 2000 seconds). We used the 5.1 version of the Off-Line Scientific Analysis (OSA) \citep{gold} of the software. The spectral analysis was done with the XSPEC package v.\ 11.3 and a 0.02 systematic error was added to the data.

\section{Time behaviour}

Light curves of all the monitoring period with IBIS in three energy bands, 20--40, 40--60 and 60--120 keV have been extracted.
In the three bottom panel of Figure 1 are shown the IBIS ligth curves. Each cross corresponds to  the source counts rate for a time bin of 1 SCW at a signal to noise ratio grater than 3$\sigma$. 

\begin{figure}[ht]
\centering
\includegraphics[height=9cm,angle=90]{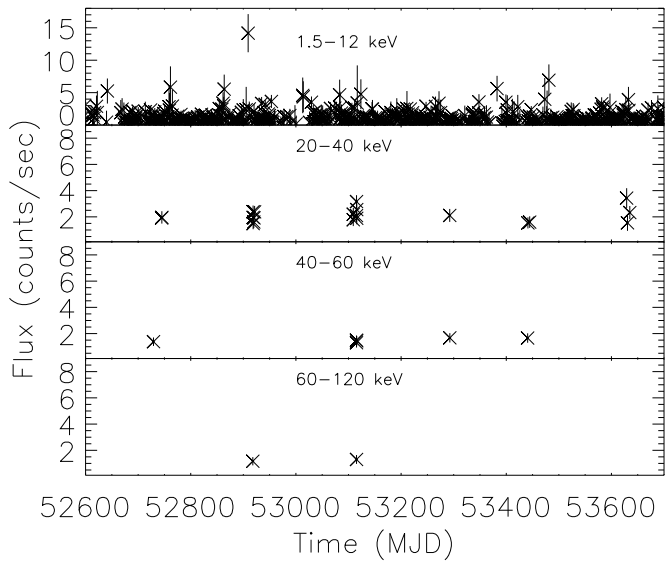}
\caption{2003-2005 ligth curve of XB 1832--330 with IBIS (with time bin of 1 SCW) and ASM (with time bin of 1 day). Each point rapresent the count rate to time bin. \label{fig1}}
\end{figure}

In the top panel of Fig. \ref{fig1} it is also reported the  {\it RXTE\/}/ASM\footnote{public data available at http://xte.mit.edu/ASM$\_$lc.html.} one day averaged light curve in the 1.5--12 keV for the same observation period. 
The source shows a weak variability in all the energy bands. 
No hint of spectral transition was detected. The corresponding flux in the 20--40 keV is of about 19 mCrab.

\begin{figure}[ht]
\centering
\includegraphics[height=5.7cm]{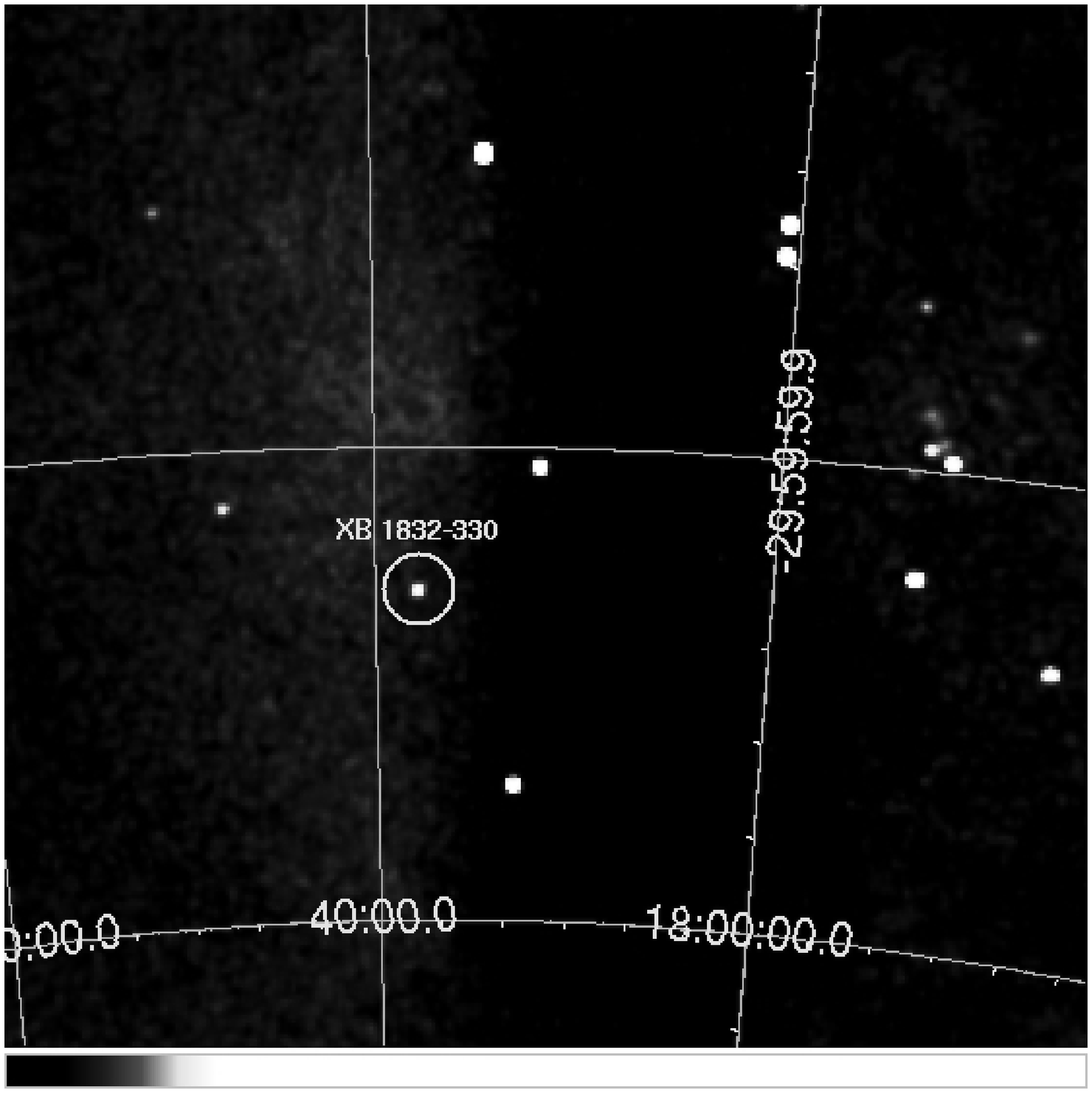}
\includegraphics[height=5.4cm]{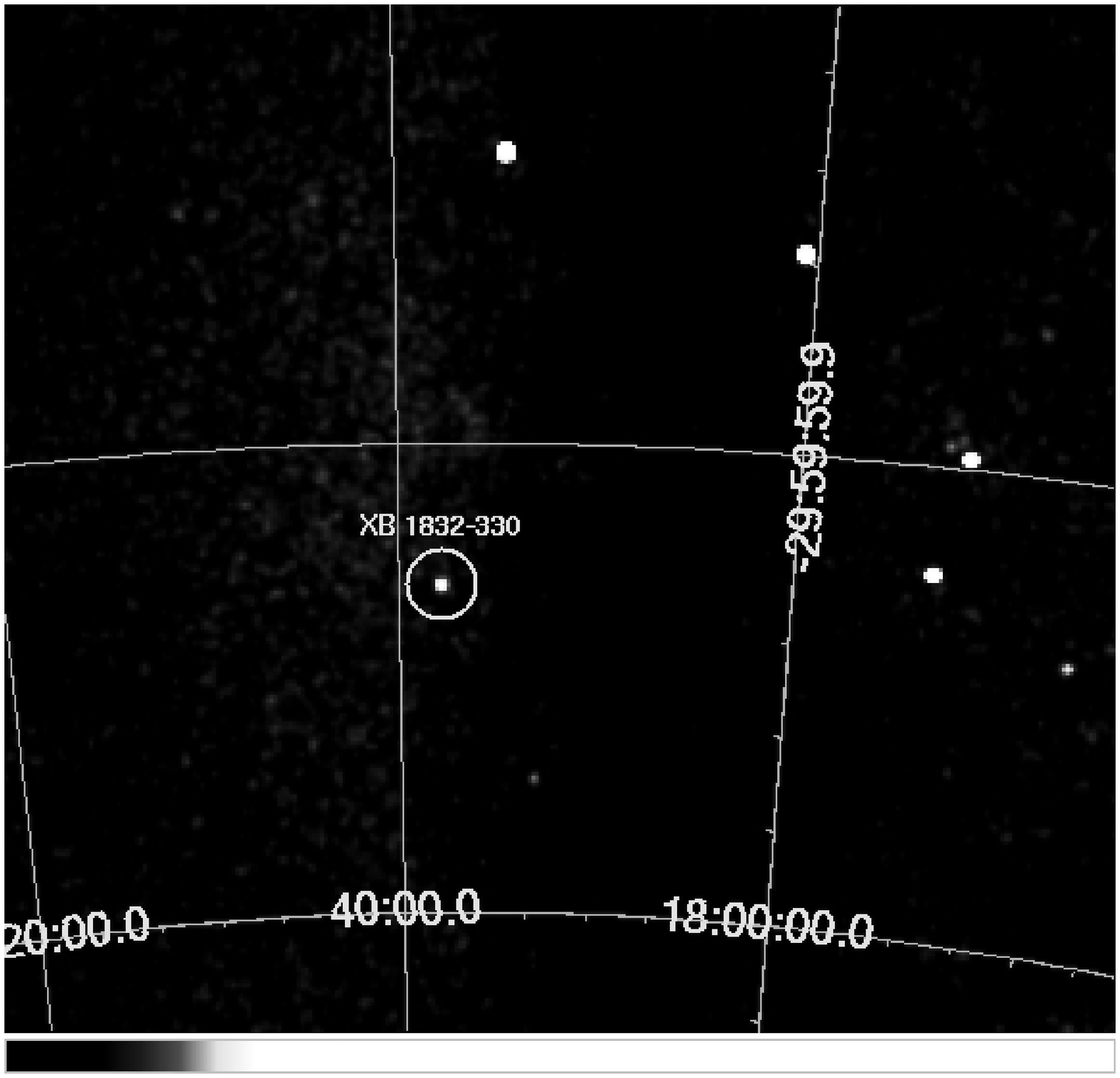}
\caption{IBIS 20--40 keV (top) and 40--60 keV (bottom) image of XB 1832--330. \label{fig2}}
\end{figure}

In Fig. \ref{fig2} are shown the IBIS mosaic image lasting 250 ks in the 20--40 and 40--60 KeV, with J2000 coordinate grid system. 

\section{Spectral analisys}

We have extracted spectra for each pointing and because of the lack of flux variation, as revealed by the IBIS (and ASM) light curves, we added all the available spectra to increase the signal to noise ratio. The final spectrum correspond to a total exposure time of 178 ks.

\begin{figure}[ht]
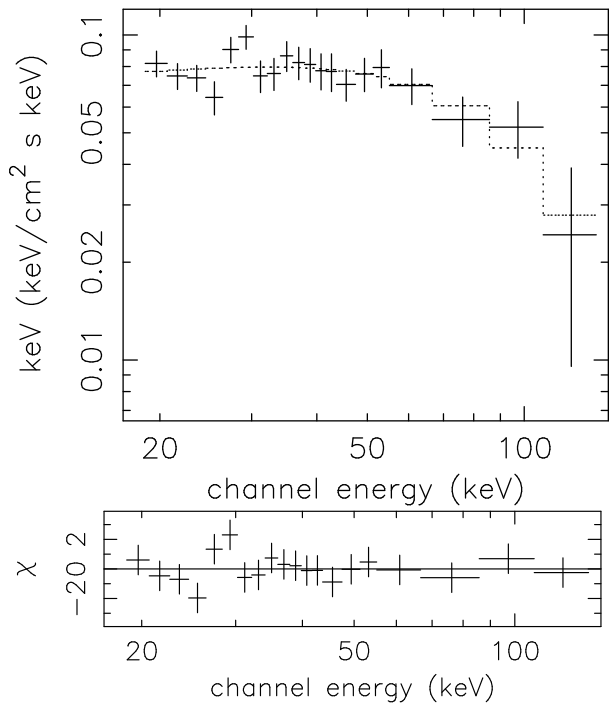

\centering
\includegraphics[height=8.0cm,angle=-90]{fig_spe_eeuf.ps}
\includegraphics[height=7.8cm,angle=-90]{fig_spedel.ps}
\caption{IBIS everage spectrum of XB 1832--330. The model (dotted line) is the CompTT comptonization model. See Table 1 for the best fit parameters. The residuals to the model are shown in the bottom panel.\label{fig3}}
\end{figure}
\begin{figure}[ht]
\centering
\includegraphics[height=8.0cm,angle=-90]{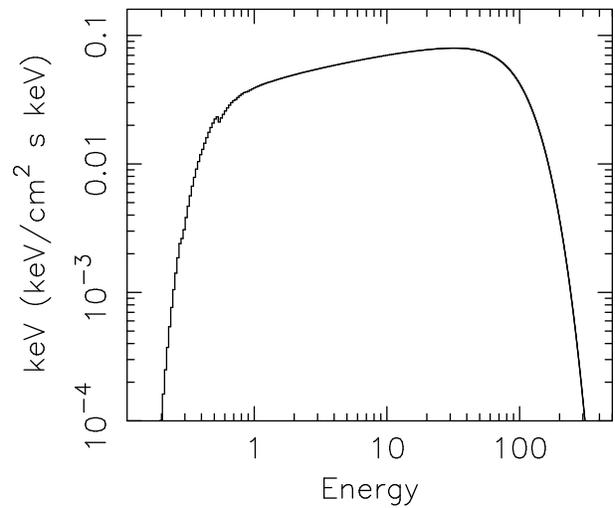}
\caption{Comptonization model corrected by absorbtion for the spectrum of XB 1832-330. \label{fig4}}
\end{figure}

This spectrum extends up to 150 keV and it is well described by a Comptonization model (CompTT) \citep{t} with a thermal plasma temperature, $kT_{\rm e}$, of about 22 keV 
 and optical depth of the electrons plasma, $\tau$, of about 2, with a $\chi_\nu^2$=0.9.
The temperature of the seed photons, $kT_{\rm 0}$, has been frozen at the best fit value of 0.15 keV, because this photons are well below the IBIS energy band.

\begin{table}[b]
  \begin{center}
    \caption{Best fit parameters of CompTT model for the IBIS data of XB 1832-330.}\vspace{1em}
    \renewcommand{\arraystretch}{1.2}
    \begin{tabular}[h]{lc}
      \hline
      Parameter          & Value\\
\hline
   $kT_{0}$ (keV)        & 0.15 (frozen) \\
   $kT_{\rm e}$ (keV)    & 21.3$^{+12.8}_{-4.7}$ \\
   $\tau$                & 1.8$^{+0.7}_{-0.9}$ \\
   norm$_{\rm CompTT}$   &  0.006$^{+0.001}_{-0.002}$  \\
   $\chi_\nu^2$(d.o.f)   &  0.9(17)\\
\hline
   $F_{\rm 20-40 keV}$ (erg s$^{-1}$ cm$^{-2}$) &8.8$\times 10^{-11}$ \\
   $F_{\rm 40-100 keV}$ (erg s$^{-1}$ cm$^{-2}$)&9.7$\times 10^{-11}$  \\
   $L_{\rm bol}$ (erg s$^{-1}$)&5.5$\times 10^{36}$  \\
          \hline \\
      \end{tabular}
    \label{tab:table}
  \end{center}
\end{table}
Data and model of the spectrum are reported in Fig. \ref{fig3}, while in Table 1 are reported the best fit parameters, as well the flux in different energy bands and the bolometric luminosity estimated for a source distance of 9.2 kpc. In Fig. \ref{fig4} is reported the best fit model of comptonization (corrected additon of the interstellar absorbtion, N$_{H}$, of 4.6$\times 10^{20}$ atom cm$^{-2}$ \citep{p}).

The electron plasma temperature and the plasma optical depth are in the typical range values of the LMXBs when in hard spectral state.

\section{Discussion and conclusion}

The {\it INTEGRAL\/}/IBIS long monitoring of the source XB 1832--330  shows that the spectrum of this compact source extend up to 150 keV and it is well described by a thermal comptonization with the temperature of the electrons, $kT_{\rm e}$, of 22 keV and optical depth, $\tau$, of 1.8. These parameter values are compatible with those found in previous BeppoSAX analisys ($kT_{\rm e}=25.3\pm1.8$ keV $\tau=1.77\pm 0.07$ \citep{p}). Moreover, assuming a distance of 9.2 kpc, the 0.1-200 keV luminosity as derived from the model, is 5.4$\times 10^{36}$ erg s$^{-1}$ that is consistent with the previous BeppoSAX measurement.

The IBIS result indicate that, in the hard X-ray band, the source is a persistent, "stable" one and spend most of the time in the typical hard state of LMXBs. This behaviour is similar to the one showed by 4U 1812--12 in the same domain of hard X-ray as reported by Tarana et al. 2006 \citep{tarana}, even if both sources are atoll and therefore changes in flux and spectral shape are expected.

The hard component of XB 1832--330 is explained to be due to the comptonization of soft photons coming from the accretion disk and the neutron star surface. 
Unluckly no data are available for the {\it INTEGRAL\/} X-ray detector, JEM-X, and then we can not investigate the presence of a disk blackbody component that indicate emission from the disk or/and the neutron star surface (as shown in \citep{p}). 

\section*{Acknowledgments}

This work has been supported by the Italian Space Agency through the grant I/R/046/04. The authors thank to Catia Spalletta for the activity support at Rome.


\end{document}